\documentclass[aps,prl,showpacs,twocolumn,superscriptaddress,floatfix,preprintnumbers,longbibliography]{revtex4-2}
\usepackage{graphicx}
\usepackage{dcolumn}
\usepackage{bm}
\usepackage{braket}
\usepackage{hyperref}
\usepackage{amsmath}
\usepackage{amsthm,amssymb}
\usepackage{mathrsfs}
\usepackage{color}
\usepackage{mathtools}
\let\OLDthebibliography\thebibliography
\renewcommand\thebibliography[1]{
  \OLDthebibliography{#1}
  \setlength{\parskip}{0pt}
  \setlength{\itemsep}{0pt plus 0.1ex}
}
\hypersetup{colorlinks=true, linkcolor=blue, anchorcolor=blue, urlcolor=blue, citecolor=blue}
\renewcommand{\eqref}[1]{\hyperref[#1]{(}\ref{#1}\hyperref[#1]{)}}
\newcommand{\g}{^3A_2}
\newcommand{\e}{^3E}

\begin{document}

\title{Nonlinear magnetic sensing with hybrid nitrogen-vacancy/magnon systems}

\author{Zhongqiang Hu}
\affiliation{Department of Electrical Engineering and Computer Science, Massachusetts Institute of Technology, Cambridge, MA 02139, USA}

\author{Zhiping He}
\affiliation{Department of Electrical Engineering and Computer Science, Massachusetts Institute of Technology, Cambridge, MA 02139, USA}

\author{Qiuyuan Wang}
\affiliation{Department of Electrical Engineering and Computer Science, Massachusetts Institute of Technology, Cambridge, MA 02139, USA}

\author{Chung-Tao Chou}
\affiliation{Department of Electrical Engineering and Computer Science, Massachusetts Institute of Technology, Cambridge, MA 02139, USA}
\affiliation{Department of Physics, Massachusetts Institute of Technology, Cambridge, MA 02139, USA}

\author{Justin T. Hou}
\affiliation{Department of Electrical Engineering and Computer Science, Massachusetts Institute of Technology, Cambridge, MA 02139, USA}

\author{Luqiao Liu}
\email{luqiao@mit.edu}
\affiliation{Department of Electrical Engineering and Computer Science, Massachusetts Institute of Technology, Cambridge, MA 02139, USA}

\date{\today}

\begin{abstract}

Magnetic sensing beyond linear regime could broaden the frequency range of detectable magnetic fields, which is crucial to various microwave and quantum applications. Recently, nonlinear interactions in diamond nitrogen-vacancy (NV) centers, one of the most extensively studied quantum magnetic sensors, are proposed to realize magnetic sensing across arbitrary frequencies. In this work, we enhance these capabilities by exploiting the nonlinear spin dynamics in hybrid systems of NV centers and ferri- or ferro-magnetic (FM) thin films. We study the frequency mixing effect in the hybrid NV/magnon systems, and demonstrate that the introduction of FM not only amplifies the intensity of nonlinear resonance signals that are intrinsic to NV spins, but also enables novel frequency mixings through parametric pumping and nonlinear magnon scattering effects. The discovery and understanding of the magnetic nonlinearities in hybrid NV/magnon systems position them as a prime candidate for magnetic sensing with a broad frequency range and high tunablity, particularly meaningful for nanoscale, dynamical, and non-invasive materials characterization.

\end{abstract}

\maketitle

Optically active spin defects as represented by nitrogen-vacancy (NV) centers in diamond, have been pursued as magnetometers for their fine spatial resolution and high sensitivity \cite{Rondin2014, Degen2017, Casola2018, Barry2020}. These defect spins are typically utilized for sensing magnetic fields either at very low frequencies (up to a few MHz) or near their intrinsic transition frequency \cite{Taylor2008, Balasubramanian2008, Acosta2009, Maertz2010, Wolf2015, Aslam2017, Lovchinsky2017, Ku2020, Wangh2022}. Recently, nonlinear spin dynamics in NV centers has been discovered \cite{Wang2022, Hu2024}, where spin transitions are excited by the sum or difference of two or more applied microwaves, providing opportunities for wide-band field sensing through frequency mixings. Meanwhile, interesting high-order magnonic dynamics has also been detected in magnetic materials proximate to NV spins, illustrating rich nonlinear phenomena such as high harmonic generation \cite{Koerner2022} and frequency comb of magnon modes \cite{Carmiggelt2023}. In these works, the nonlinear scattering within the magnonic system alone is proposed to account for the observed frequency mixings, and NV spins are treated as purely linear magnetic sensors that optically read out the mixed signals. The parallel discoveries of nonlinear resonances in the two closely related systems bring in intriguing questions: where indeed does the nonlinearity happen in the NV/magnon system? Do the interactions among magnons play the dominant role or do magnons act as simple field amplifiers that boost the nonlinear dynamics inside NV spins? The absence of a clear understanding on the contributions from the two spin-based systems impedes the optimization of these nonlinear effects, and  can lead to controversies in experimental study. For example, the magnon-scattering picture encounters difficulties in explaining the observed mixing effects when driven frequencies are within the magnon band gap \cite{Carmiggelt2023}.

To account for the interactive dynamics in the hybrid system that involves microwave photons, NV spins, and magnons, we carry out a systematic study on the nonlinear interactions in hybrid NV/ferri- or ferro-magnet (FM) systems, where FM = $\mathrm{Y_3Fe_5O_{12}}$ (YIG) or $\mathrm{Ni_{80}Fe_{20}}$ (NiFe) thin films. By quantitatively comparing the resonance spectra between NV/magnon and NV-only systems, we find that the nonlinear spin dynamics in both NV centers and magnetic thin films contributes to the nonlinear magnetic resonances observed in the hybrid system. The introduction of magnons influences the optically detected magnetic resonances (ODMR) in two aspects: firstly it enhances the strength of nonlinear NV spin transitions by amplifying the microwave fields, and meanwhile it also provides more efficient routes for frequency mixings through nonlinear magnon scatterings. Moreover, the nonlinear magnonic dynamics enables new transitions that are forbidden in the NV-only system such as frequency mixings involving fractional integer coefficients. Our comparison of different magnetic thin films – YIG and NiFe also suggests the important role played by nanoscale domain structures in breaking the translational symmetry for magnon scattering. The comprehensive understanding of frequency mixings in NV/magnon systems not only provides a sensitive probe for studying nonlinear magnonics covering a broad frequency and wave-vector range, but also presents a novel platform for general quantum magnetic sensing.

\begin{figure}[t!]
\centering	\includegraphics[width=1\columnwidth]{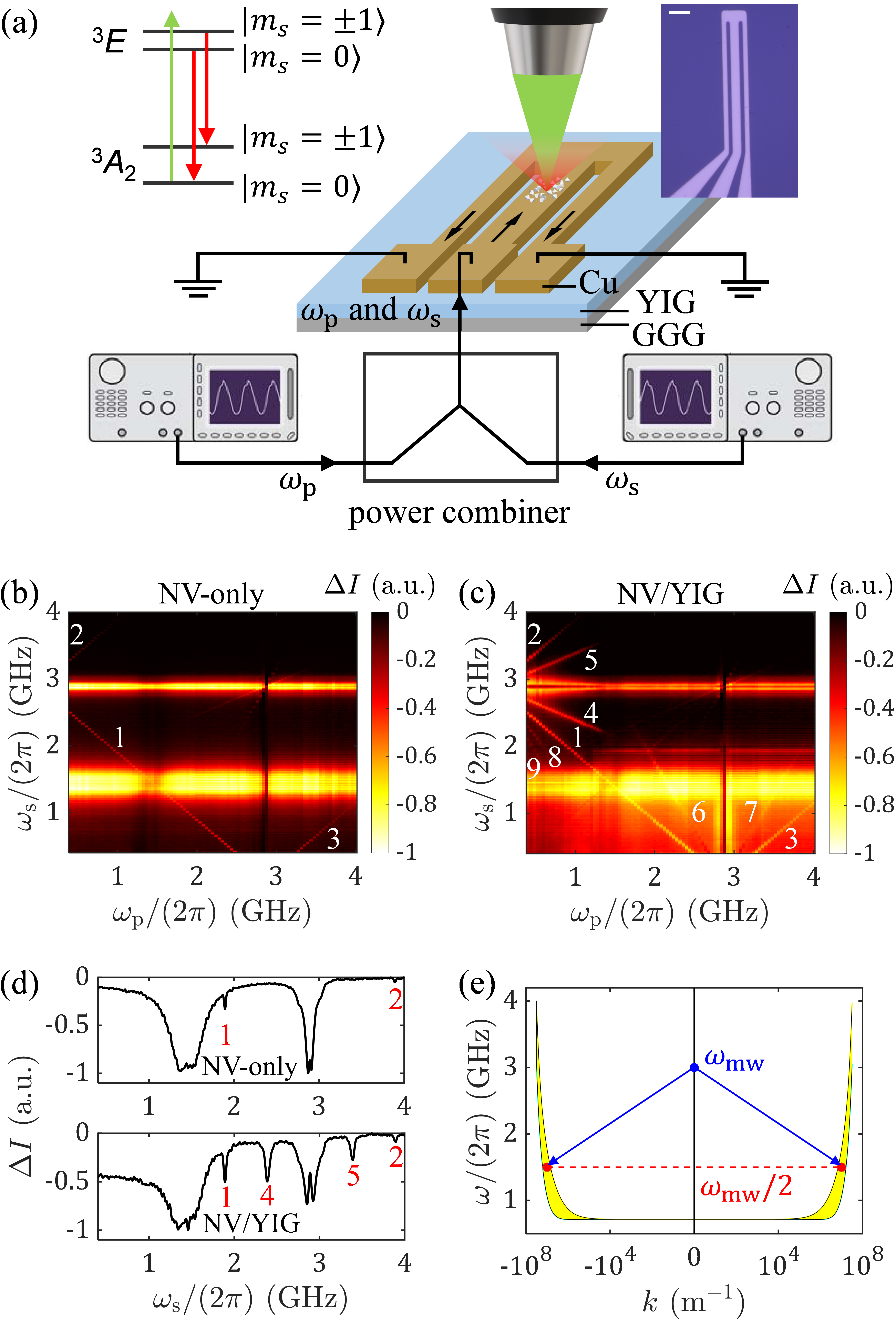}
\caption{(a) Experimental scheme for local, nonlinear magnetic sensing in a hybrid NV/YIG system. Top-left inset: energy level diagram of an NV spin. Top-right inset: photograph of the real device with a scale bar of $10 \ \mathrm{\mu m}$. (b) Resonance spectra of an NV-only system under $\omega_\mathrm{p,s}$ with applied microwave powers $P_\mathrm{p,s} = 1.5 \ \mathrm{mW}$. (c) Resonance spectra of the NV/YIG system with $P_\mathrm{p,s} = 1.5 \ \mathrm{mW}$. (d) Top and bottom show $\omega_\mathrm{s}$ scan in (b) and (c) when $\omega_\mathrm{p} / (2\pi) = 1 \ \mathrm{GHz}$. (e) Illustration of parametric pumping process inside YIG. The magnon band continuum (yellow region) is calculated for $40$-nm-thick YIG with an in-plane external field of 100 Oe.}
\label{fig1}
\end{figure}

Figure \hyperref[fig1]{1(a)} illustrates the experimental scheme of a hybrid NV/YIG sensing system. To study the nonlinear spin dynamics, microwaves from two independent signal generators are combined through a power combiner, which is confirmed to not induce any frequency mixing effect. The microwaves are further applied onto a copper antenna with a 4-$\mathrm{\mu m}$-wide signal line, on top of a 40-$\mathrm{nm}$-thick YIG thin film. Microdiamond particles with an average diameter of $\sim 750 \ \rm  nm$ and an NV concentration of $\sim 3.5 \ \rm ppm$ are dispersed on top of the antenna. The energy levels of an NV center is shown in the top-left inset of Fig. \hyperref[fig1]{1(a)}. The optical ground state $\g$ and excited state $\e$ are spin triplets with spin sublevels separated by $\omega_\mathrm{A} / (2\pi) = 2.87 \ \mathrm{GHz}$ and $\omega_\mathrm{E} / (2\pi) = 1.42 \ \mathrm{GHz}$, respectively, under zero external field \cite{Doherty2012, Rogers2009}. Microwaves at the sublevel splitting $\omega_\mathrm{A,E}$ can induce linear transitions from $\ket{m_s=0}$ to $\ket{m_s=\pm1}$, which strengthens the non-radiative optical transition and yields a reduction of photoluminescence (PL) intensity excited by 532 nm green laser \cite{Acosta2010, Dreau2011, Choi2012, Jensen2013}.

Beyond the linear regime, multiple microwave photons can also excite NV spin transitions when their frequency sum or difference matches the transition frequency. As we demonstrated recently \cite{Hu2024}, this originates from the nonlinear wave-spin interactions when the microwave magnetic fields have components both along and perpendicular to the principle axes of NV spins. Figure \hyperref[fig1]{1(b)} illustrates typical resonance spectra when NV spins are driven under two microwaves: pump $\omega_\mathrm{p}$ and signal $\omega_\mathrm{s}$. Here the antenna is directly fabricated on a $\mathrm{Gd_3Ga_5O_{12}}$ (GGG) substrate. Aside from the strong, wide linear resonance at $\omega_\mathrm{s} = \omega_\mathrm{A,E}$, additional bright lines emerge at $\omega_\mathrm{s} + \omega_\mathrm{p} = \omega_\mathrm{A}$ (labelled as 1) and $\omega_\mathrm{s} - \omega_\mathrm{p} = \pm\omega_\mathrm{A}$ (labelled as 2 and 3). Note that in our experiment, in order to enhance the signal-to-noise ratio, we modulate the amplitude of the $\omega_\mathrm{s}$ input with a low frequency of 17.77 Hz and measure the change of PL $\Delta I$ using a lock-in amplifier. This selective modulation has led to the asymmetric features in Fig. \hyperref[fig1]{1(b)} along $\omega_\mathrm{p}$ and $\omega_\mathrm{s}$ scans. 

\begin{figure}[t]
\centering	\includegraphics[width=1\columnwidth]{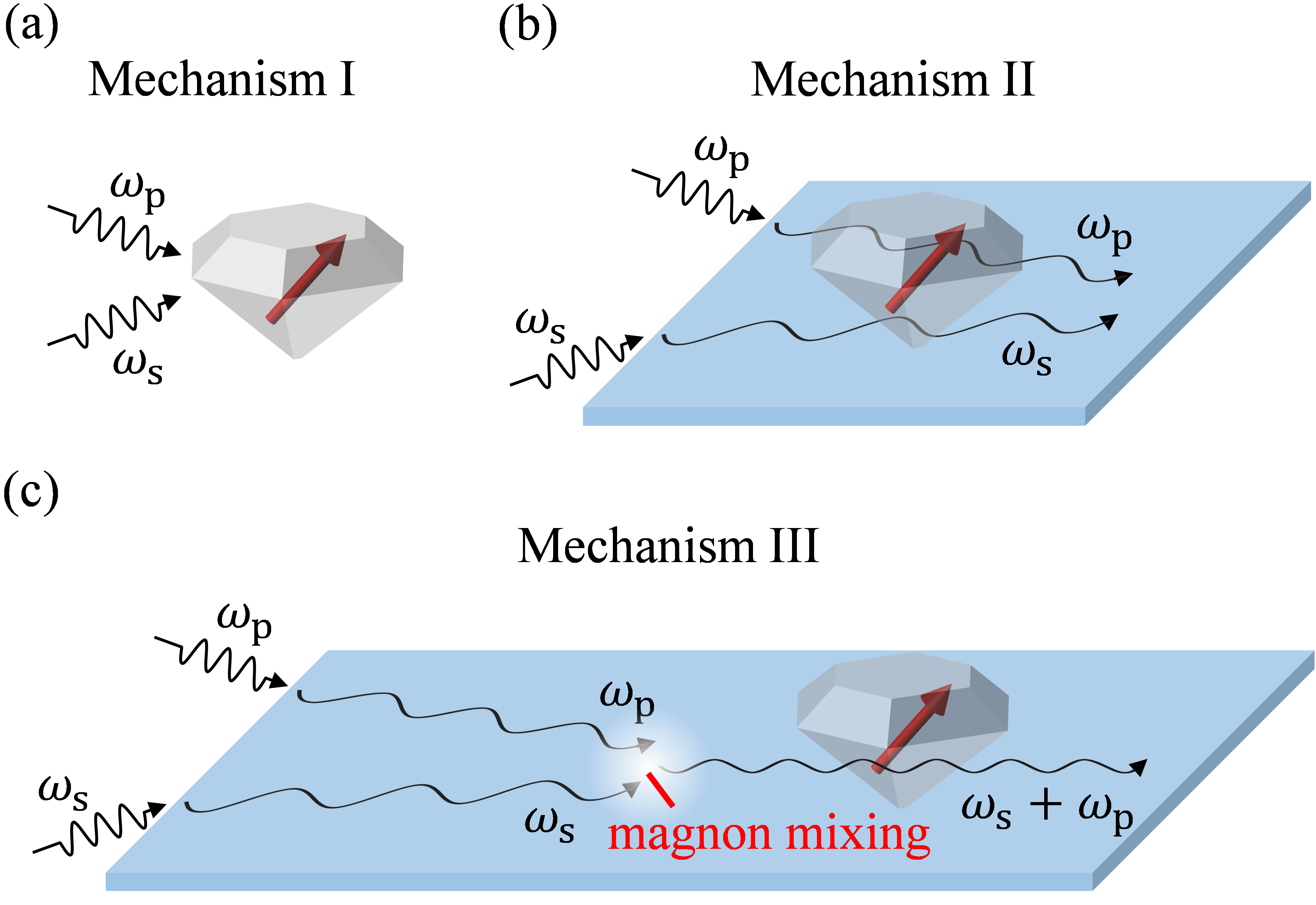}
\caption{Three possible mechanisms for the nonlinear resonance at $\omega_\mathrm{s} + \omega_\mathrm{p} = \omega_\mathrm{A}$ in NV/YIG. (a) Microwave photons at $\omega_\mathrm{p,s}$ directly excite nonlinear NV spin transitions. (b) Magnons at $\omega_\mathrm{p,s}$ are linearly generated and their stray fields excite the nonlinear dynamics of NV spins. In both (a) and (b), the nonlinearity happens inside NV spins. (c) The nonlinear magnon scattering generates magnons at $\omega_\mathrm{s} + \omega_\mathrm{p} = \omega_\mathrm{A}$ within YIG, which excite the NV spin transitions through linear dipolar interactions.}
\label{fig2}
\end{figure}

In Fig. \hyperref[fig1]{1(c)}, we show $\Delta I$ versus $\omega_\mathrm{p,s}$ for the hybrid NV/YIG system. Examples of $\omega_\mathrm{s}$ scans under fixed $\omega_\mathrm{p}/(2\pi) = 1 \ \mathrm{GHz}$ are illustrated in Fig. \hyperref[fig1]{1(d)} for NV-only and NV/YIG separately. We notice that besides the lowest-order nonlinear resonances of 1-3, which have been observed in the NV-only system, additional resonance signals show up in NV/YIG, at various linear combinations of $\omega_\mathrm{p,s}$, labelled as 4-9: (4) $\omega_\mathrm{s} + \omega_\mathrm{p} / 2 = \omega_\mathrm{A}$, (5) $\omega_\mathrm{s} - \omega_\mathrm{p} / 2 = \omega_\mathrm{A}$, (6) $\omega_\mathrm{p} + \omega_\mathrm{s} / 2 = \omega_\mathrm{A}$, (7) $\omega_\mathrm{p} - \omega_\mathrm{s} / 2 = \omega_\mathrm{A}$, (8) $\omega_\mathrm{s} + 3\omega_\mathrm{p} / 2 = \omega_\mathrm{A}$, and (9) $\omega_\mathrm{s} + 2\omega_\mathrm{p} = \omega_\mathrm{A}$. Firstly, the frequency mixings involving one or multiple halves of applied microwave frequencies have not been observed before, and can be explained by the parametric pumping process of magnons \cite{Damon1953, Suhl1957}, where one $\omega_\mathrm{p,s}$ microwave photon generates two $\omega_\mathrm{p,s}/2$ magnons with the same frequency and opposite wave vectors [Fig. \hyperref[fig1]{1(e)}]. This parametric process benefits from the ellipticity of excited magnons due to the dipolar interaction which is not present in the NV-only system. The parametrically generated $\omega_\mathrm{p,s}/2$ magnons, along with the linearly excited ones or microwave photons themselves at $\omega_\mathrm{p,s}$, can further mix together and lead to resonances in NV spins. Specifically, resonances labeled as 4-7 can be explained with this consecutive picture involving two magnons or photons in the latter step, while resonances 8-9 need to involve three magnons/photons, corresponding to higher-order dynamics.

While it is clear that the emergence of new resonance signals involving halves of microwave frequencies have benefited from parametrically pumped magnons, the detailed mechanisms behind the frequency mixings of two or more signal sources require additional examinations, considering that both NV spins themselves and magnonic thin films can exhibit nonlinear dynamics. Taking the sum frequency resonance at $\omega_\mathrm{s} + \omega_\mathrm{p} = \omega_\mathrm{A}$ as an example, Figs. \hyperref[fig2]{2(a)-2(c)} illustrate three possible mechanisms. Mechanism I [Fig. \hyperref[fig2]{2(a)}] represents the simplest case, where two microwave photons directly excite the spin transition inside NV centers. With the addition of YIG thin film, magnons at $\omega_\mathrm{p,s}$ can be linearly generated by the corresponding microwave photons and their stray fields can excite the nonlinear dynamics of NV spins [Mechanism II in Fig. \hyperref[fig2]{2(b)}]. Under this scenario, the nonlinearity still happens inside NV spins while magnon excitations effectively amplify the oscillating fields. On the other hand, the scattering of $\omega_\mathrm{p,s}$ magnons could also lead to the generation of magnons at $\omega_\mathrm{s} + \omega_\mathrm{p} = \omega_\mathrm{A}$ within YIG [Mechanism III in Fig. \hyperref[fig2]{2(c)}]. These nonlinearly generated magnons can further influence the NV spin transitions through linear dipolar interactions.

\begin{figure}[t]
\centering	\includegraphics[width=1\columnwidth]{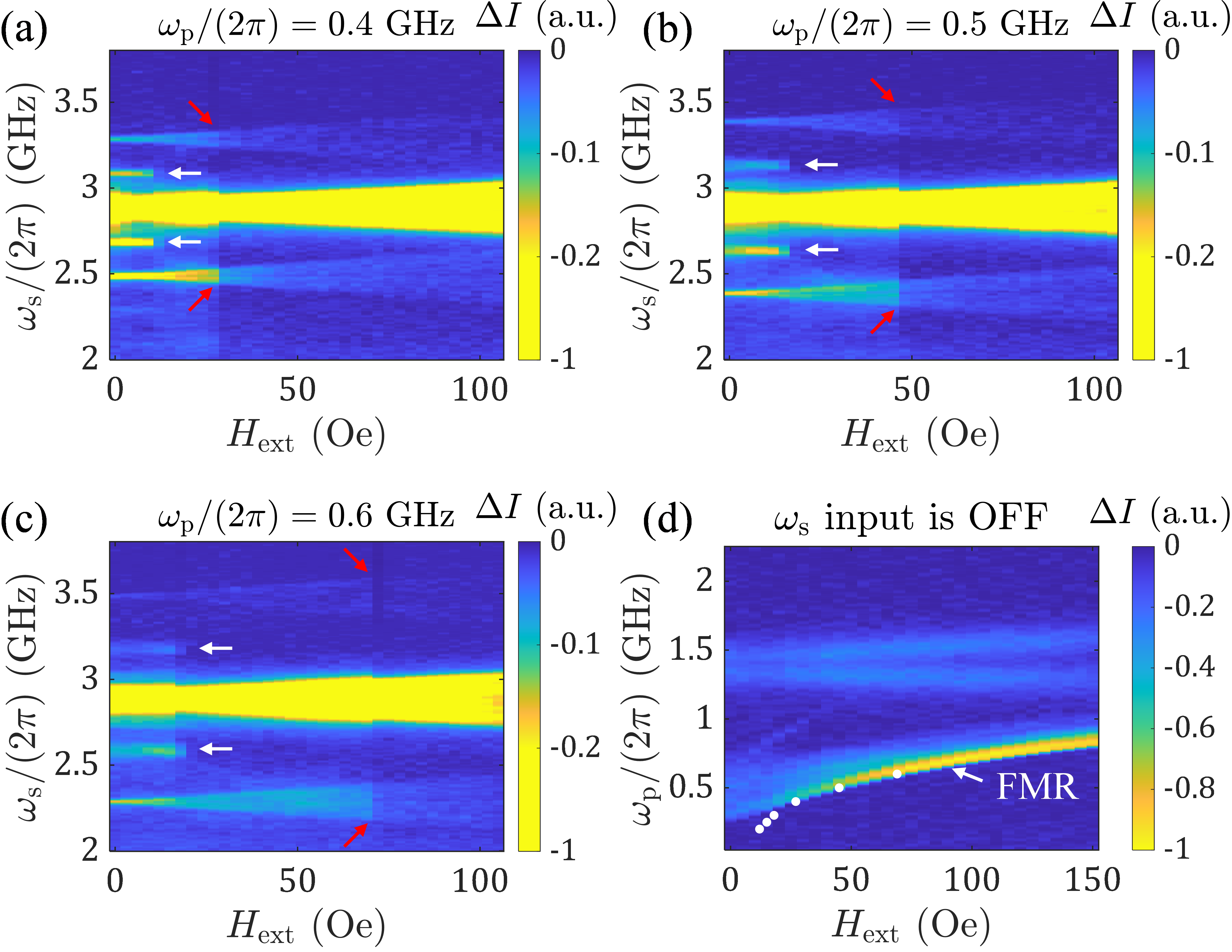}
\caption{(a)-(c) $\Delta I$ versus $\omega_\mathrm{s}$ and $H_\mathrm{ext}$ with $P_\mathrm{p,s} = 0.3 \ {\rm mW}$ for NV/YIG, when $\omega_\mathrm{p}/(2\pi)$ is fixed at (a) $0.4 \ {\rm GHz}$, (b) $0.5 \ {\rm GHz}$, and (c) $0.6 \ {\rm GHz}$. $H_\mathrm{ext}$ is in-plane and forms an $45^\circ$ angle with the microwave currents. (d) $\Delta I$ versus $\omega_\mathrm{p}$ and $H_\mathrm{ext}$ with $P_\mathrm{s} = 0$ and $P_\mathrm{p} = 0.1 \ {\rm mW}$. The white dots correspond to critical fields where the minimum of magnon band reaches $0.2 \ \mathrm{GHz}$, $0.25 \ \mathrm{GHz}$, $0.3 \ \mathrm{GHz}$, $0.4 \ \mathrm{GHz}$, $0.5 \ \mathrm{GHz}$, and $0.6 \ \mathrm{GHz}$ (from left to right).}
\label{fig3}
\end{figure}

To evaluate magnons' contribution to the nonlinear resonance signals, we vary the external magnetic field applied onto the hybrid system $H_\mathrm{ext}$, effectively altering the position of magnon band. By shifting the magnon band via $H_\mathrm{ext}$, we can compare the resonance signals when the driving microwave frequency is above and below the minimum magnon frequency. In the latter scenario, no magnons should be excited. Figures \hyperref[fig3]{3(a)-3(c)} illustrate $\Delta I$ versus $\omega_\mathrm{s}$ and $H_\mathrm{ext}$ when the pump frequency is fixed at $\omega_\mathrm{p}/(2\pi) = 0.4 \ {\rm GHz}$, $0.5 \ {\rm GHz}$, and $0.6 \ {\rm GHz}$, respectively. We can clearly identify the resonances near $\omega_\mathrm{s} \pm \omega_\mathrm{p} = \omega_\mathrm{A}$ (red arrows) and $\omega_\mathrm{s} \pm \omega_\mathrm{p} /2 = \omega_\mathrm{A}$ (white arrows), which exhibit cone-shaped structure as $H_\mathrm{ext}$ varies. The broadening with $H_\mathrm{ext}$ is similar to the linear resonance of $\omega_\mathrm{s} = \omega_\mathrm{A}$, as a result of the random orientations of NV center principle axes in the diamond particles. We notice that the resonances of $\omega_\mathrm{s} \pm \omega_\mathrm{p} /2 = \omega_\mathrm{A}$ suddenly disappear when $H_\mathrm{ext}$ is increased to certain values [12 Oe, 15 Oe, and 18 Oe, respectively in Figs. \hyperref[fig3]{3(a)-3(c)}]. Meanwhile, the resonance strengths at $\omega_\mathrm{s} \pm \omega_\mathrm{p} = \omega_\mathrm{A}$ also experience abrupt drops to lower values when $H_\mathrm{ext}$ reaches 27 Oe, 45 Oe, and 69 Oe, respectively. To correlate the observed cut-off fields, we measure the frequency of magnons with zero wave vectors, i.e., ferromagnetic resonance (FMR) mode \cite{Wolfe2014, McCullian2020, Lee2020}, by applying only the $\omega_\mathrm{p}$ source [Fig. \hyperref[fig3]{3(d)}]. The six pairs of cut-off fields and corresponding minimum magnon frequencies extracted from Figs. \hyperref[fig3]{3(a)-3(c)} are labelled as white dots in Fig. \hyperref[fig3]{3(d)}, which are consistent with the FMR curve. Here, we have taken into account that the parametrically pumped magnons are actually at $\omega_\mathrm{p}/2$. The small deviation at the low frequency end in Fig. \hyperref[fig3]{3(d)} may be related to the discrepancy between the FMR frequency and the minimum magnon frequency \cite{Rezende2020}. The existence of cut-off frequencies (fields) and their consistency with the minimum of magnon band confirms the magnon contribution in the nonlinear resonance signals via mechanism II or III. Meanwhile, the reduced but finite resonance signals at $\omega_\mathrm{s} \pm \omega_\mathrm{p} = \omega_\mathrm{A}$ beyond the cut-off fields [red arrows in Figs. \hyperref[fig3]{3(a)-3(c)}] suggest that in the absence of magnon scattering, the NV nonlinearity can still lead to finite resonances via mechanism I or II. Previously, nonlinear resonances in NV/magnon systems have been attributed to the magnon scattering alone \cite{Carmiggelt2023}. Our verification on the finite contribution from the NV nonlinearity therefore helps to reconcile the controversy between the observed finite nonlinear resonance signals and the fact that the driven frequencies are in the magnon band gap.

\begin{figure}[t]
\centering	\includegraphics[width=1\columnwidth]{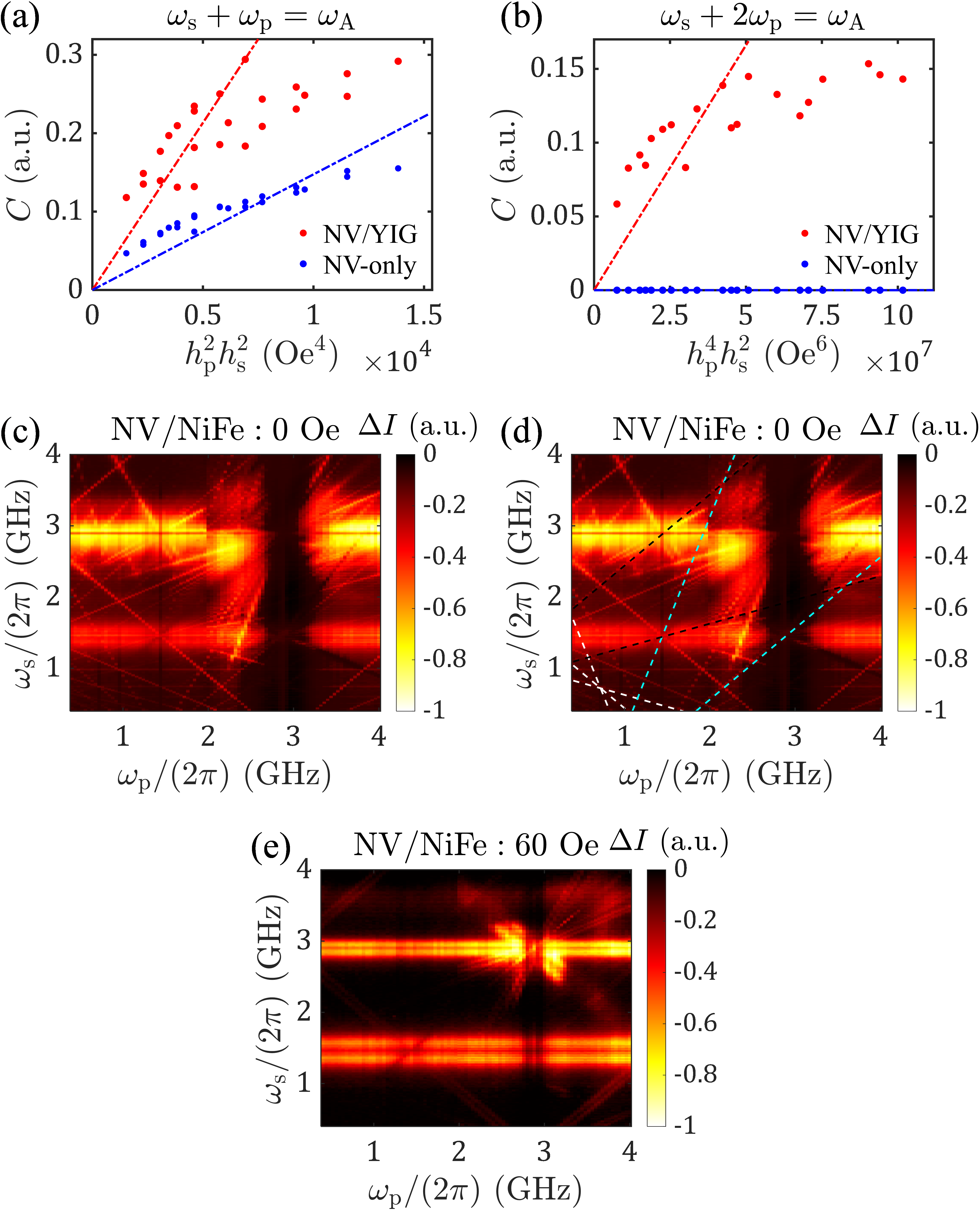}
\caption{(a) Strength $C$ of the $\omega_\mathrm{s} + \omega_\mathrm{p} = \omega_\mathrm{A}$ resonance with $\omega_\mathrm{p} / (2\pi) = 0.6 \ \mathrm{GHz}$ for NV/YIG (red dots) and NV-only (blue dots) systems as a function of $h_\mathrm{p}^2h_\mathrm{s}^2$. (b) Strength $C$ of the $\omega_\mathrm{s} + 2\omega_\mathrm{p} = \omega_\mathrm{A}$ resonance with $\omega_\mathrm{p} / (2\pi) = 0.3 \ \mathrm{GHz}$ for NV/YIG (red dots) and NV-only (blue dots) systems as a function of $h_\mathrm{p}^4h_\mathrm{s}^2$. (c) Resonance spectra of an NV/NiFe system under $\omega_\mathrm{p,s}$ with $P_\mathrm{p,s} = 8 \ \mathrm{mW}$ and zero external field. (d) Same as (c), highlighting resonances at $\omega_\mathrm{s} + 3\omega_\mathrm{p} = \omega_\mathrm{A}$, $2\omega_\mathrm{s} + 2\omega_\mathrm{p} = \omega_\mathrm{A}$, $3\omega_\mathrm{s} + \omega_\mathrm{p} = \omega_\mathrm{A}$ (white, dashed lines), $2\omega_\mathrm{p} - 2\omega_\mathrm{s} = \omega_\mathrm{A}$, $3\omega_\mathrm{p} - \omega_\mathrm{s} = \omega_\mathrm{A}$ (blue lines), $2\omega_\mathrm{s} - 2\omega_\mathrm{p} = \omega_\mathrm{A}$, and $3\omega_\mathrm{s} - \omega_\mathrm{p} = \omega_\mathrm{A}$ (black lines). (g) Resonance spectra of the NV/NiFe system with $H_\mathrm{ext} =$ 60 Oe.}
\label{fig4}
\end{figure}

While the field-dependent experiment gives a clear evidence on the important role of magnons, the direct source of the nonlinearity remains unclear, noticing that magnons contribute to both Mechanisms II and III. To get a quantitative understanding, we carry out measurements and analysis on the power dependence of nonlinear resonance signals. When the NV nonlinearity is the only source of frequency mixings, the strength $C$ of the resonance signals at $\omega_\mathrm{s} + \omega_\mathrm{p} = \omega_\mathrm{A}$ can be calculated using basic parameters of diamond \cite{SM}, which is shown to be proportional to $h_\mathrm{s}^2h_\mathrm{p}^2$, where $h_\mathrm{s,p}$ is the magnetic field strength of corresponding microwave. By calibrating the microwave field on top surface of the antenna and varying the applied microwave powers, we obtain the relationship between $C$ and $h_\mathrm{p}^2h_\mathrm{s}^2$ [blue dots in Fig. \hyperref[fig4]{4(a)}], in agreement with the expected power dependence (blue, dashed curve). When YIG is present, the magnons’ contribution from Mechanism II [Fig. \hyperref[fig2]{2(b)}] enters through their amplification on the microwave fields, where excited magnons around the antenna lead to higher $h_\mathrm{s,p}$ than the NV-only case through their stray fields. Considering that the amplified fields strengthen both the linear and nonlinear resonances in the NV/YIG system, we can therefore calibrate $h_\mathrm{s,p}$ using the linear resonance signals at $\omega_\mathrm{s} = \omega_\mathrm{A}$ \cite{SM}, with which the stray fields of linearly generated magnons get accounted for. The red dots in Fig. \hyperref[fig4]{4(a)} illustrate $C$ under calibrated $h_\mathrm{p,s}$ for the NV/YIG system. We notice that under the same, calibrated $h_\mathrm{p,s}$, $C$ is always higher for the NV/YIG system compared to the NV-only system, suggesting that the field amplification effect (Mechanism II) is not enough to explain the enhancement of the nonlinear resonance signals, and Mechanism III must have played an important role. In Fig. \hyperref[fig4]{4(b)}, an even stronger contrast between the NV-only and NV/YIG systems is found on the third-order resonance at $\omega_\mathrm{s} + 2\omega_\mathrm{p} = \omega_\mathrm{A}$ [line 9 in Fig. \hyperref[fig1]{1(c)}], where the resonance is only observed when YIG is present, demonstrating that the generation of $\omega_\mathrm{s} + 2\omega_\mathrm{p}$ magnons is the main source of the resonance. In Figs. \hyperref[fig4]{4(a)} and \hyperref[fig4]{4(b)}, we notice that $C$ of NV/YIG deviates from the expected linear dependence on $h_\mathrm{p}^2h_\mathrm{s}^2$ and $h_\mathrm{p}^4h_\mathrm{s}^2$ at
the high field end, which may reflect the saturation of
nonlinear magnon scattering at high powers.

Besides the energy conservation as reflected in $\omega_\mathrm{s} \pm \omega_\mathrm{p} = \omega_\mathrm{A}$, etc., nonlinear magnon scatterings also require the conversation of momentum, which is not always easily satisfied for an arbitrary set of magnon frequencies due to their complicated dispersion relationship. We note that the breaking of translation symmetry in the fabricated devices due to dimension confinement through patterned antenna and thin-film geometry helps to release this constraint. To this end, NiFe represents another candidate material where nonlinear magnon scattering happens more readily due to the breaking of the whole film into nanoscale domain structures from the stronger magnetic anisotropy \cite{Koerner2022, Rable2023}. Figure \hyperref[fig4]{4(c)} shows $\Delta I$ versus $\omega_\mathrm{p,s}$ under zero external field for an NV/NiFe system, where even richer resonance curves are observed compared to NV/YIG. In Fig. \hyperref[fig4]{4(d)}, we label fourth-order resonance curves of $m\omega_\mathrm{s} + n\omega_\mathrm{p} = \omega_\mathrm{A}$ with integers $m$ and $n$ satisfying $|m| + |n| = 4$, which are not observed in the NV/YIG system. The importance of the domain structure in the NiFe film is confirmed with the application of a higher external field. Figure \hyperref[fig4]{4(e)} shows the resonance spectra under $H_\mathrm{ext}$ = 60 Oe, where the high-order signals are greatly suppressed.

\begin{figure}[t]
\centering	\includegraphics[width=1\columnwidth]{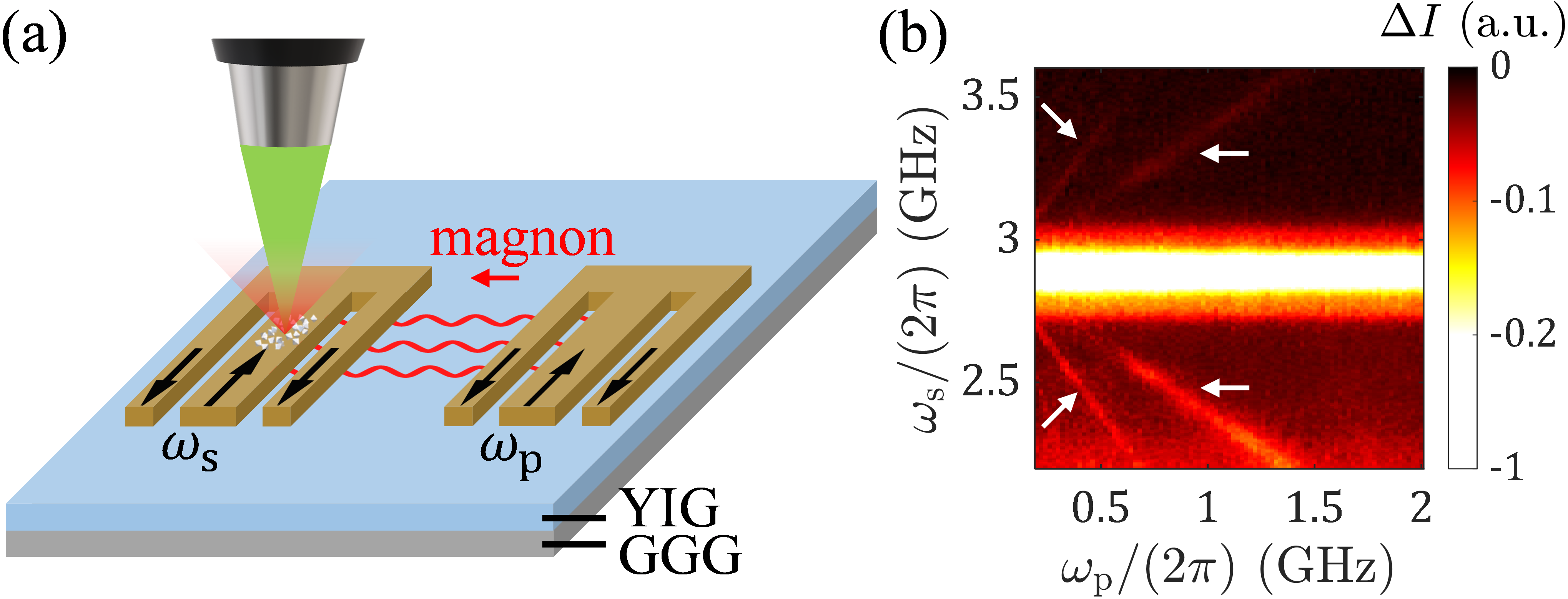}
\caption{(a) Schematic of experimental configuration for detecting nonlocally excited magnons. (b) $\Delta I$ versus $\omega_\mathrm{p,s}$ with $P_\mathrm{p,s} = 1.5 \ \mathrm{mW}$ and $H_\mathrm{ext}$ 10 Oe.}
\label{fig5}
\end{figure}

A direct application of the nonlinear resonances in NV/magnon systems is to employ them as a magnon spectrometer, irrespective of whether the magnon excitation and detection happen locally with the same antenna or they are separated far apart. As an example, in Fig. \hyperref[fig5]{5(a)} we apply $\omega_\mathrm{p,s}$ microwaves to two antennae separated by $18 \ \mathrm{\mu m}$, on top of a 40-$\mathrm{nm}$-thick YIG film. The linearly or parametrically excited magnons from the $\omega_\mathrm{p}$ microwave are transmitted to the other antenna and detected by the $\omega_\mathrm{s}$ microwave. Figure \hyperref[fig5]{5(b)} illustrate $\Delta I$ versus $\omega_\mathrm{p,s}$ when $H_\mathrm{ext} = 10 \ \mathrm{Oe}$ is applied in plane. The four resonance curves at $\omega_\mathrm{s} \pm \omega_\mathrm{p} = \omega_\mathrm{A}$ and $\omega_\mathrm{s} \pm \omega_\mathrm{p} / 2 = \omega_\mathrm{A}$ (white arrows) can be clearly observed.

In conclusion, we study the frequency mixings in hybrid NV/magnon systems. We demonstrate that both the nonlinearity in NV spins and the nonlinear magnonic dynamics contribute to the resonances at mixed frequencies. Compared with the intrinsic nonlinear effects in NV spins, the introduction of FM enables new mixings at half integers of driving frequencies, as a result of the parametric pumping effect. Meanwhile, the interactions between magnons and NV spins also enrich the nonlinear resonance signals through linear field amplification and nonlinear magnon scattering effects, where the latter one is facilitated by the broken translational symmetry from the device fabrication and nanoscale domain structures. The nonlinear interactions in hybrid NV/magnon systems promise novel functionalities for quantum control and sensing. Particularly, one can leverage the frequency mixing effect to extract the spectrum information of spin waves over a broad wave-vector range. Meanwhile, the hybrid NV/magnon system can also be used as a whole for the detection of free space magnetic fields at arbitrary microwave frequencies.

\begin{acknowledgments}
The work is supported by the National Science Foundation under award DMR-2104912. Z.H. and C.-T.C. acknowledge support from Semiconductor Research Corporation (SRC) and DARPA. Q.W. acknowledges support from the MathWorks Fellowship.
\end{acknowledgments}

\bibliography{main_text}

\end{document}